\documentclass{ws-p9-75x6-50}
\usepackage{graphicx,amsmath,amssymb,bm,multicol}

\begin{document}

\title{Effective Nucleon-Nucleon Interaction and\\ 
Fermi Liquid Theory}
\author{Achim Schwenk, Gerald E. Brown}
\address{Department of Physics and Astronomy,
State University of New York,\\Stony Brook, N.Y. 11794-3800,
E-mail: aschwenk@nuclear.physics.sunysb.edu}
\author{Bengt Friman}
\address{Gesellschaft f\"ur Schwerionenforschung, Planckstr. 1, 64291
Darmstadt, Germany}


\maketitle

\abstracts{We present two novel relations between the
quasiparticle interaction in nuclear matter and the unique low
momentum nucleon-nucleon interaction in vacuum.
These relations provide two independent constraints on the Fermi
liquid parameters of nuclear matter. Moreover, the new constraints
define two combinations of Fermi liquid parameters, which are 
invariant under the renormalization group flow in the
particle-hole channels. Using empirical values for the
spin-independent Fermi liquid parameters, we are able to 
compute the major spin-dependent ones by imposing the new 
constraints as well as the Pauli principle sum rules.}

The work we present at this conference~\cite{FLTVlowk}
was motivated by the results of
Bogner \etal~\cite{Vlowk}, who have constructed a low momentum
nucleon-nucleon interaction $V_{\text{low k}}$ using traditional 
nuclear effective interaction methods. Starting from 
a realistic nucleon-nucleon interaction, such as the Paris, Bonn-A,
and Argonne potentials or a chiral effective field theory model, 
they integrate out relative momenta larger than a cutoff $\Lambda$
in the sense of the renormalization group (RG). The hard momenta 
renormalize $V_{\text{low k}}$, such that the original low momentum 
half-on-shell $T$ matrix, i.e. phase shifts and low momentum
components of the scattering wave functions, as well as 
bound state properties remain unchanged. 

Diagrammatically $V_{\text{low k}}$ sums all ladders with bare
potential vertices and intermediate momenta greater than the
cutoff. Subsequently, the energy dependence of the ladder sum is
removed by adding the folded diagram corrections. For the case of the
nonrelativistic two-nucleon problem, the nuclear model space
effective interaction methods are equivalent to solving a RG flow 
equation~\cite{VlowkRG} obtained by requiring $d \, T(k^{\prime},k;k^2) /
d \Lambda  = 0$ with the various bare potentials as large $\Lambda$
initial conditions.
\begin{figure}
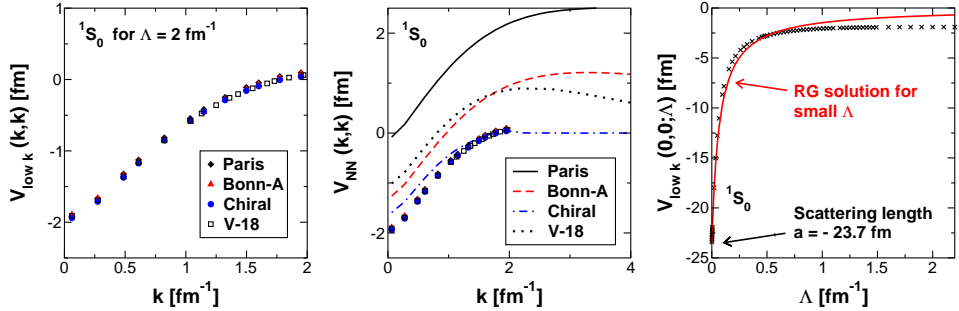

\begin{center}
\includegraphics[scale=0.22,clip=]{p4vlowkdiag2.0.eps}
\hspace{0.05cm}
\includegraphics[scale=0.22,clip=]{pshift.eps}
\hspace{0.05cm}
\includegraphics[scale=0.22,clip=]{p1s0lambda.eps}
\end{center}
\caption{Diagonal elements of $V_{\text{low k}}$ for different bare
potentials in the $^1 S_0$ channel at a cutoff $\Lambda = 2.0 \;
\text{fm}^{-1}$ and the comparison to the bare potentials
$V_{\text{NN}}$. RG flow of $V_{\text{low k}} (0,0;\Lambda)$ 
versus $\Lambda$ in the $^1 S_0$ channel. Figures taken from Bogner
\etal~\cite{Vlowk}.}
\label{fig:Vlowk}
\end{figure}
The results of the RG decimation to $V_{\text{low k}}$ are shown in
Fig.~(\ref{fig:Vlowk}). For $\Lambda \lesssim 2 \; \text{fm}^{-1}$ all 
model potentials flow to a {\it unique} low momentum interaction 
$V_{\text{low k}}$, which to a good approximation and for reasonable
values of the cutoff is merely shifted by a constant compared to the bare
potentials. This constant shift in momentum space, corresponding to a 
smeared delta function in coordinate space, is of such strength that 
it removes the experimentally undetermined core from the bare models.
For $\Lambda > m_\pi$, $V_{\text{low k}}$ can be projected on pion
exchange terms, which determine the low momentum potential shape, plus 
contact terms, which account for the short range part of the
interaction. In this setup, the non-pionic contributions of the bare
potentials flow to ``fixed point'' values. 

In accordance with the ideas of effective field theories, the results 
are insensitive to the cutoff, provided there is a separation of scales.
The insensitivity can best be seen in the plateau of the RG flow of 
$V_{\text{low k}} (0,0;\Lambda)$, Fig.~(\ref{fig:Vlowk}). The lowest order
approximation to $V_{\text{low k}}(0,0)$ is given by the effective
energy dependent potential $V_{\text{eff}}(0,0;p^{2}=0)$ at zero
energy. For small $\Lambda$, the RG equation for energy
dependent effective potentials of Birse \etal~\cite{Birse}
is solved by $V_{\text{eff}}(0,0;p^2=0) = (1/a - 2 \Lambda /\pi)^{-1}$.
This is plotted as solid line in Fig.~(\ref{fig:Vlowk}) and agrees
well with $V_{\text{low k}}$.

In this work we will take $V_{\text{low k}}$ as dynamical input to 
Fermi liquid theory. This results in an intriguing relation between 
the effective interactions in vacuum and in medium. It is even more so,
as the role of the local repulsive core in the bare interactions is taken 
over by non-local parts, which in turn contribute to the non-locality 
of the quasiparticle interaction in Fermi liquid theory. For $\Lambda
\sim k_{\text{F}}$ the energy independent $V_{\text{low k}}$ includes the
effects of the repulsive core, is generally smooth, and can
therefore be used as a $G$ matrix. This is approximate from a 
Brueckner theory standpoint, since self-energy insertions and 
the dependence on the center of mass momentum are ignored. 
However, it is expected that the effects of
the self-energy insertions are small. The novel relations
have as input the s-wave matrix elements $V_{\text{low k}}(0,0)$ at 
$\Lambda = k_{\text{F}}$, i.e. in the weakly $\Lambda$ dependent
region. Similar energy independent
effective potentials are in agreement with $V_{\text{low k}}$ up to a
few percent~\cite{UCOM,Okubo}. As the Fermi liquid
parameters are fixed points under the RG flow towards the Fermi
surface, the constraints relate $V_{\text{low k}}$ to these
fixed points.

At low temperatures, strongly interacting normal Fermi systems can be 
described by weakly interacting quasiparticles and quasiholes.
As in any effective theory, the quasiparticle interaction is only restricted by
the symmetries. For nuclear matter these are rotations in space, spin, and 
isospin, and the quasiparticle interaction ${\mathcal
F}$ in units of the density of states at the Fermi surface is
given by
\begin{equation}
{\mathcal F} = \sum_{l} \biggl( F_l + F_l^{\prime} \: 
{\bm \tau} \cdot {\bm \tau^{\prime}} + G_l \: {\bm \sigma} \cdot {\bm 
\sigma^{\prime}} + G^{\prime}_l \: {\bm \tau} \cdot {\bm 
\tau^{\prime}} \: {\bm \sigma} \cdot {\bm \sigma^{\prime}} \biggr) P_l 
(\cos \theta) + \mbox{tensor} + {\mathcal O}(A^{-1/3}) ,
\end{equation}
where $\theta$ denotes the angle between the quasiparticles ${\mathbf
p}$ and ${\mathbf p^\prime}$ on the Fermi surface. The Fermi liquid 
parameters of this effective theory are determined by experiment.
On a microscopic level, the quasiparticle interaction includes
all diagrams that are quasiparticle-quasihole irreducible in the zero
sound channel.

In order to satisfy the Pauli principle, one needs an integral
equation, which generates an infinite set of diagrams for the quasiparticle
interaction. This is achieved by the induced interaction of Babu and
Brown~\cite{BB,BBfornuclmat}. The induced interaction in combination
with the antisymmetrized bare interaction as driving term generates 
the complete particle-hole parquet, i.e. all fermionic planar diagrams without 
the particle-particle channel~\cite{parquet}. 
This is the minimum set of diagrams mandated by the
Pauli principle. The induced interaction incorporates the
response of the system to the presence of the quasiparticle.
We approximate the driving term, which in principle includes all particle-hole
irreducible diagrams, e.g. particle-particle ladders and non-planar
diagrams, by $z^2 V_{\text{low k}}$ with a density dependent 
cutoff $\Lambda = k_{\text{F}}$. Given the driving term, the 
induced interaction is {\it
exact} for ${\mathbf p} = {\mathbf p^{\prime}}$. Solving the integral
equations for the scattering amplitude and the induced interaction in
this limit simultaneously leads to the two constraints, in units where
the nucleon mass $m=1$
\begin{align}
& \sum_{l} \biggl\{ 2 F_l - \frac{F_l}{1+F_l/(2l+1)} + 2
F_l^{\prime}  - \frac{F_l^{\prime}}{1+F_l^{\prime}/(2l+1)}  - 3 
\bigl( 2 G_l - \frac{G_l}{1+G_l/(2l+1)} \bigr) \nonumber \\
& \hspace{0.1cm} - 3  \bigl( 2 G_l^{\prime} -
\frac{G_l^{\prime}}{1+G_l^{\prime}/(2l+1)} \bigr) \biggr\}  =  z^2 
\, \frac{16  k_{\text{F}} (1+F_1/3)}{\pi} \, V_{\text{low k}}
(0,0;\Lambda=k_{\text{F}},{}^1\text{S}_0) 
\label{C1} \\
& \sum_{l} \biggl\{ 2 F_l - \frac{F_l}{1+F_l/(2l+1)}  -3  \bigl( 2
F_l^{\prime}  - \frac{F_l^{\prime}}{1+F_l^{\prime}/(2l+1)} \bigr)  
+  2 G_l - \frac{G_l}{1+G_l/(2l+1)} \nonumber \\
& \hspace{0.1cm} -3  \bigl( 2 G_l^{\prime} 
- \frac{G_l^{\prime}}{1+G_l^{\prime}/(2l+1)} \bigr)
\biggr\}  =  z^2 \, \frac{16  k_{\text{F}}
(1+F_1/3)}{\pi} \, V_{\text{low k}} (0,0;\Lambda=k_{\text{F}},{}^3\text{S}_1) .
\label{C2}
\end{align}

In analogy to Fermi liquid theory, where one separates the
quasiparticle contribution of the full Green's function from the
multi-pair background, one can separate the soft modes of the
quasiparticle-quasihole propagators from the hard ones. We show that
in this setup and neglecting the flow in the BCS channel, 
the two constraints as well as the two Pauli principle sum rules are
RG invariant under the flow to the Fermi surface.~\cite{FLTVlowk} 
The induced interaction permits the
calculation of the beta function for the quasiparticle scattering
amplitude and interaction to all orders in the limit ${\mathbf
p} \to {\mathbf p^\prime}$ and thus considerably improves one-loop
approximations thereof. To lowest order, our RG equations agree with
the perturbative one-loop flow of Dupuis~\cite{Dupuis}.

\begin{figure}
\begin{center}
\includegraphics[scale=0.23,clip=]{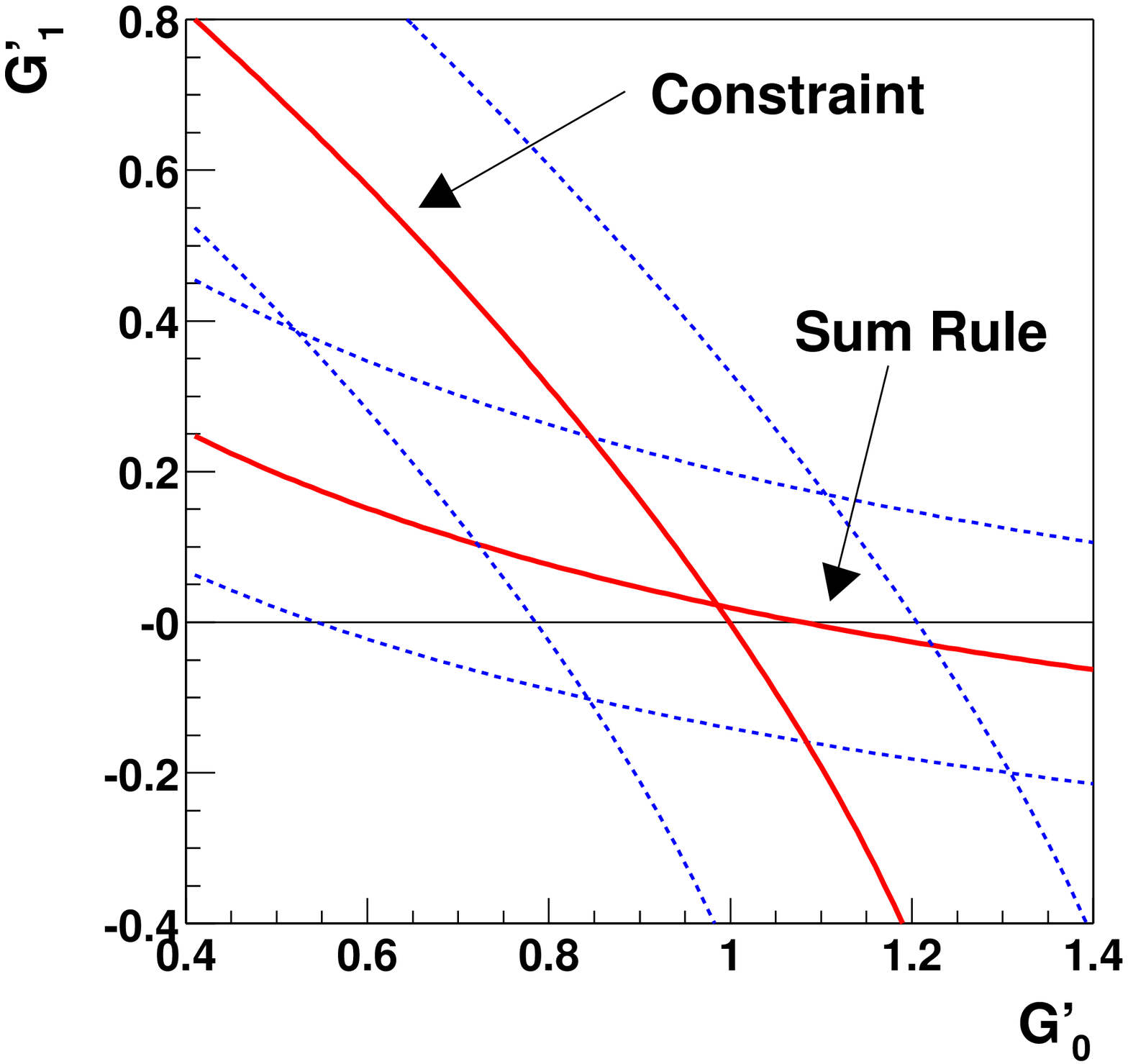}
\hspace{0.5cm}
\includegraphics[scale=0.23,clip=]{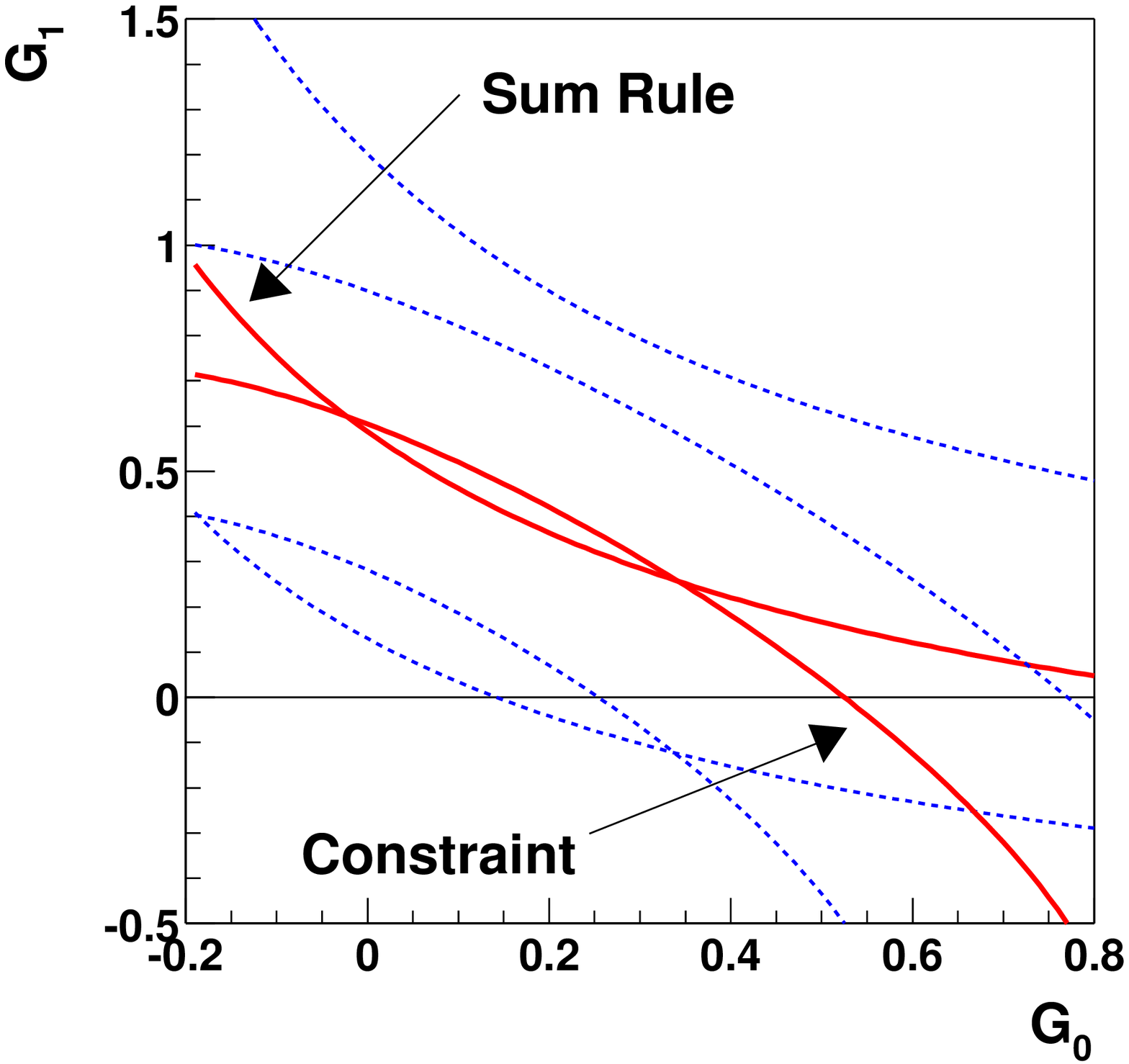}
\caption{The solution for the spin-dependent Fermi liquid parameters
(solid lines) with error bands limited by the dashed lines. Here
the effect of tensor parameters is neglected.}
\label{fig:fitnotensor}
\end{center}
\end{figure}

Are these constraints consistent with phenomenological values for the
Fermi liquid parameters? For this purpose, we approximate the
Legendre series with its $l = 0,1$ terms. We use
phenomenological values for the spin-independent parameters and then
compute the spin dependent ones imposing the Pauli principle sum rules
as well as the constraints. From the incompressibility, the effective
mass, the symmetry energy, the anomalous orbital gyromagnetic ratio,
and a self-consistent calculation of the nucleon spectral function,
one extracts
\begin{align}
F_0 = -0.27 \quad F_1 = -0.85 \quad F_0^{\prime} = 0.71 \quad
F_1^{\prime} = 0.14 \quad z = 0.8 .
\end{align}
In Fig.~(\ref{fig:fitnotensor}) we show the solution for the
spin-dependent Fermi liquid parameters without tensor interactions.
We thus find
\begin{align}
G_0 = 0.15 \pm 0.3 \quad G_1 = 0.45 \pm 0.3 \quad
G_0^\prime = 1.0 \pm 0.2 \quad G_1^\prime = 0 \pm 0.2 .
\end{align}
The value of $G_0^\prime$ is to be compared to the experimental
constraint from the Gamow-Teller resonance, after correcting for the
effects of the Delta-hole polarization $G_0^\prime = 1.0$.
Tensor interactions are included easily into the constraints.
A simple argument however shows that the tensor parameters have 
to be treated self-consistently within the induced 
interaction~\cite{FLTVlowk}. 
The one-bubble recoupling of $G_0^\prime$ to tensor interactions 
reduces the tensor parameters significantly, in agreement with the
results of Dickhoff \etal~\cite{Dickhoff}. 

A solution of the flow equations, as presented by Metzner for the
Hubbard model at this conference, would provide the scattering 
amplitude also for non-forward scattering, which is
of high interest for the calculation of superfluid gaps and
transport processes, e.g. in neutron star interiors.

\section*{Acknowledgments}
We thank Scott Bogner and Tom Kuo for helpful discussions. AS
thanks the Theory Group at GSI for their kind hospitality. This
work was supported by the US-DOE grant No. DE-FG02-88ER40388.


\begin{thebibliography}{99}
\bibitem{FLTVlowk} A. Schwenk, G.E. Brown, B. Friman, nucl-th/0109059,
submitted to NPA.
\bibitem{Vlowk} S.K. Bogner \etal, nucl-th/9912056, nucl-th/0108041.
\bibitem{VlowkRG} S.K. Bogner, A. Schwenk, T.T.S. Kuo, G.E. Brown, 
to be published (2001).
\bibitem{Birse} M.C. Birse \etal, \Journal{\PLB}{464}{169}{1999}.
\bibitem{UCOM} T. Neff, Ph.D. Thesis in preparation, Darmstadt 
University, (2001).
\bibitem{Okubo} E. Epelbaoum \etal, \Journal{\NPA}{645}{413}{1999}.
\bibitem{BB} S. Babu and G.E. Brown, \Journal{\AP}{78}{1}{1973}.
\bibitem{BBfornuclmat} O. Sj\"oberg, \Journal{\AP}{78}{39}{1973}.
\bibitem{parquet} A. Lande and R.A. Smith, \Journal{\PRA}{45}{913}{1992}.
\bibitem{Dupuis} N. Dupuis, cond-mat/9604189.
\bibitem{Dickhoff} W.H. Dickhoff \etal, \Journal{\NPA}{405}{534}{1983}.
\end{thebibliography}
\end{document}